\documentclass[10pt]{blois07}
\usepackage{graphicx}
\usepackage{cite,./mcite}
\usepackage{ccaption}
\captiondelim{ }

\begin{document}
\title{Blois07/EDS07\\ Proceedings}

\author{M. M. Islam$^{a}$, J. Ka$\check{s}$par$^{b}$ and R. J. Luddy$^{c}$}

\institute{
$^{a}$\footnotesize{Department of Physics, University of Connecticut, Storrs, CT 06269, USA, (islam@phys.uconn.edu)}\\ 
$^{b}$ TOTEM Collaboration, CERN, Geneva, Switzerland (Jan.Kaspar@cern.ch)\\
$\;\;\;$Institute of Physics, Academy of Sciences of the Czech Republic, Prague\\
$^{c}$ Department of Physics, University of Connecticut, Storrs, CT 06269, USA, (RJLuddy@phys.uconn.edu)\\
}

\maketitle{\bf{pp Elastic Scattering at LHC in a Nucleon--Structure Model}}

\it{(Presented by M. M. Islam)}\em

\begin{abstract}
We predict pp elastic differential cross sections at LHC at c.m.\ energy 14 TeV 
and momentum transfer range $\vert t \vert$ = 0 -- 10 GeV$^2$ in a nucleon-structure model.  In 
this model, the nucleon has an outer cloud of quark-antiquark condensed ground state, 
an inner shell of topological baryonic charge ($r\simeq 0.44 F$) probed by the vector 
meson $\mathrm{\omega}$, and a central quark-bag ($r\simeq 0.2 F$) containing valence quarks.  
We also predict $\rm{d}\sigma/d\it{t}$ in the Coulomb-hadronic interference region.  Large 
$\vert t\vert$ elastic scattering in this model arises from valence quark-quark scattering, which 
is taken to be due to the hard-pomeron (BFKL pomeron with next to leading order 
corrections).  We present results of taking into account multiple hard-pomeron 
exchanges, i.e.\ unitarity corrections.  Finally, we compare our prediction of pp 
elastic $\rm{d}\sigma/d\it{t}$ at LHC with the predictions of various other models.  Precise 
measurement of pp $\;\rm{d}\sigma/d\it{t}$ at LHC by the TOTEM group in the $\vert t\vert$ 
region 0 -- 5 GeV$^2$ 
will be able to distinguish between these models.
\end{abstract}

%

High energy pp and $\mathrm{\bar{p}p}$ elastic scattering have been at the 
forefront of accelerator research since the early seventies with the 
advent of CERN Intersecting Storage Rings (ISR) and measurement of pp 
elastic differential cross section in the c.m.\ energy range $\sqrt{s}$ = 23 -- 62 
GeV and momentum transfer range $\vert t \vert$ = 0.8 -- 10 GeV$^2$ \cite{Nagy1}. 
This was followed by the Fermilab accelerator where pp elastic scattering 
was measured at c.m.\ energy $\sqrt{s}$ = 27.4 GeV in a fixed target experiment 
at large momentum transfers: $\vert t \vert$ = 5.5 -- 14 GeV$^2$ \cite{Faiss2}.
Next came the CERN SPS Collider, where $\mathrm{\bar{p}p}$ elastic scattering was 
measured at c.m.\ energies 546 GeV and 630 GeV -- a jump of one order of 
magnitude in c.m.\ energy from ISR \cite{Bozzo3,Bozzo3a,Bernard4}. 
The Fermilab Tevatron followed next where $\mathrm{\bar{p}p}$ elastic scattering was measured at 
c.m.\ energy 1.8 TeV, but in a rather small momentum transfer range: 
$\vert t \vert$ = 0 -- 0.5 GeV$^2$ \cite{Amos5,Abe6}.  We are now at the 
threshold of a new period of accelerator research with the LHC starting 
up soon and with the planned measurement of pp elastic scattering by 
the TOTEM group at c.m.\ energy 14 TeV and momentum transfer range 
$\vert t \vert \simeq \;$0 -- 10 GeV$^2$ \cite{TOTEM7,TOTEM8}.

My collaborators and I have been studying pp and $\mathrm{\bar{p}p}$ elastic 
scattering since late seventies. From our phenomenological investigation, 
we have arrived at two results: 1) a physical picture of the nucleon, 2) an 
effective field theory model underlying the physical picture \cite{Islam9}.  The 
physical picture shows that the nucleon has an outer cloud, an inner shell 
of baryonic charge, and a central quark-bag containing the valence quarks (Fig.\ 1). 
The radius of the shell is about 0.44 F and that of the quark-bag is 0.2 F.  The 
underlying field theory model turns out to be a gauged Gell--Mann--Levy 
linear $\sigma$-model with spontaneous breakdown of chiral symmetry and with 
a Wess--Zumino--Witten (WZW) anomalous action.  The model attributes the 
outer nucleon cloud to a quark--antiquark condensed ground state analogous 
to the BCS ground state in superconductivity-- an idea that was first 
proposed by Nambu and Jona-Lasinio.  The WZW action indicates that the 
baryonic charge is geometrical or topological in nature, which is the 
basis of the Skyrmion model.  The action further shows that the vector 
meson $\mathrm{\omega}$ couples to this topological charge like a gauge boson, i.e.\ 
like an elementary vector meson.  As a consequence, one nucleon probes the 
baryonic charge of the other via $\mathrm{\omega}$-exchange.  In pp elastic 
scattering, in the small momentum transfer region, the outer cloud of one nucleon interacts with that 
of the other giving rise to diffraction scattering.  As the momentum transfer increases, 
one nucleon probes the other at intermediate distances and the $\mathrm{\omega}$-exchange becomes 
dominant.  At momentun transfers even larger, one nucleon scatters off the other via 
valence quark-quark scattering.

Our calculated pp elastic $\rm{d}\sigma/d\it{t}$ at c.m.\ energy 14 TeV is shown in Fig.\ 2 by the solid line 
that includes all three processes: diffraction, $\mathrm{\omega}$-exchange, and qq scattering.  The dotted 
curve shows $\rm{d}\sigma/d\it{t}$ due to diffraction only.  We see that diffraction dominates 
in the small $\vert t\vert$ region, but falls off rapidly.  The dot-dashed curve shows 
$\rm{d}\sigma/d\it{t}$ due to $\mathrm{\omega}$-exchange only and indicates that $\mathrm{\omega}$-exchange dominates 
in the $\vert t\vert$ region 1.5 -- 3.5 GeV$^2$.  Beyond that, the valence quark-quark scattering 
takes over.  The dashed curve for $\vert t\vert >$ 3.5 GeV$^2$ represents $\rm{d}\sigma/d\it{t}$ with 
single valence quark-quark scattering, whereas the solid curve represents $\rm{d}\sigma/d\it{t}$ with 
all multiple valence quark-quark scattering.

Let us next examine how the three processes are described in our calculations \cite{Islam9}.  
Diffraction is described by using the impact parameter representation and a phenomenological 
profile function:\vspace{-0.2cm}

$T_D (s,t)=i\,p\,W\int_0^\infty {b\;\rm{d}\it{b}\;J_0 } (b\,q)\Gamma_{D}(s,b)$;\hspace{6.61cm}                 (1)\\
$q$ is the momentum transfer ($q = \sqrt{|t|}$) and $\Gamma_{D}(s,b)$ is 
the diffraction profile function, which is related to the eikonal function $\chi_{D}(s,b)$:
$\Gamma_{D}(s,b) = 1 - \exp(i \chi_{D}(s,b))$. We take $\Gamma_{D}(s,b)$ to
be an even Fermi profile function:

$\Gamma_{D}(s,b) = g(s) [\frac{1}{1 + e^{(b-R)/a)}} +  \frac{1}{1 + e^{-(b+R)/a}} -1 ]$.\hspace{5.87cm}(2)\vspace{0.04cm}\\
The parameters $R$ and $a$ are energy dependent: $R$ = $R_0 + R_1(\ln s -$ $i\pi\over 2$ $)$,
$a$ = $a_0 + a_1(\ln s -$ $i\pi\over 2$ $)$; $g(s)$ is a complex crossing even
energy-dependent coupling strength.

The diffraction amplitude we obtain has the following asymptotic properties:\vspace{0.1cm}\\
1.  $\sigma_{\rm{tot}}(s) \sim (a_{0} + a_{1} \ln s)^{2}\hspace{2.82cm}$ (Froissart-Martin bound)\vspace{0.1cm}\\
2.  $\rho (s) \simeq \frac{\pi a_1}{a_0 + a_1 \ln s}\hspace{4.13cm}$ (derivative dispersion relation)\vspace{0.06cm}\\
3.  $T_{D}(s,t) \sim i\; s\; \ln^{2}s\; f(|t|\; \ln^{2}s)\hspace{1.72cm}$  (Auberson-Kinoshita-Martin scaling)\vspace{0.1cm}\\
4.  $T_{D}^{\mathrm{\bar{p}p}}(s,t) = T_{D}^{\mathrm{pp}}(s,t)\hspace{3.39cm}$      (crossing even)

Incidentally, the profile function (2) has been used by Frankfurt et al.\ to estimate 
the absorptive effect of soft hadronic interactions (gap survival probability) in 
the central production of Higgs at LHC \cite{Frankfurt10}.

The $\mathrm{\omega}$-exchange amplitude in our model has the form

$T_{\omega}(s,t)\;\sim \;\exp [i\;\chi _D (s,0)]\;s\;\frac{F^2(t)}{m_\omega ^2 -t}$.\hspace{7.62cm}(3)\vspace{0.1cm}\\
where the factor $s$ shows that $\mathrm{\omega}$ is behaving like an elementary spin-1 boson.  
The two form factors indicate that $\mathrm{\omega}$ is probing two baryonic charge 
distributions -- one for each nucleon.  The factor $\exp [i\;\chi _D (s,0)]$ represents 
the absorptive effect due to soft hadronic interactions.

We view large $\vert t\vert$ elastic scattering as a hard collision 
of a valence quark from one proton with a valence quark from the other proton (Fig.\ 3).  
Since this process involves high energy quark-quark scattering at large 
momentum transfer, one would expect that it should be described by perturbative 
QCD.  In fact, in perturbative QCD, the two quarks would interact via BFKL pomeron, 
that is, reggeized gluon ladders with rungs of gluons that lead to a fixed branch point 
in the angular momentum plane at $\alpha_{\rm{BFKL}}$ = 1 + $\omega$.  The value of $\omega$ in 
next to leading order lies in the range 0.13 -- 0.18 as argued by Brodsky et al.\ \cite{Brodsky11}.  
We refer to the BFKL pomeron with next to leading order corrections included as the 
``hard-pomeron".  In our calculations, we approximate the hard-pomeron by a fixed pole 
and take the corresponding qq amplitude as \cite{Islam12}

$\hat{T}_{1}(\hat{s},t)=i\; \gamma_{\mathrm{qq}}\;\hat{s}\;\left (\hat{s}\;e^{-i\frac{\pi}{2}}\right )^{\omega}
     \frac{1}{\vert t\vert +r^{-2}_{0}}$,  $\hspace{6cm}$\hspace{1.36cm}                                    (4)\\
where $\hat{s}$ is the square of the c.m.\ energy of qq scattering.

Our pp elastic $\rm{d}\sigma/d\it{t}$ calculation at 14 TeV reported earlier \cite{Islam9,Islam12} 
included only a single hard-pomeron exchange in qq scattering.  However, Eq. (4) shows 
that the hard-pomeron predicts a qq asymptotic total 
cross section $\hat{\sigma}_{\rm{tot}}(\hat{s}) \propto \hat{s}^{\omega}$, i.e.\ $\hat{\sigma}_{\rm{tot}}(\hat{s})$ 
grows like a power of $\hat{s}$ and therefore violates unitarity and the Froissart-Martin bound.  
To restore unitarity in the qq channel, we use the eikonal representation and write the full qq scattering 
amplitude as

$\hat{T}(\hat{s},t) = i\hat{p}\hat{W}\int_{0}^{\infty}b\; \rm{d}\it{b}\; J_{0}(b q)
     \left [1-e^{i\hat{\chi}(\hat{s},b)}\right ]$.\hspace{6.06cm}                                                 (5)\\
Taking $\hat{T_1}(\hat{s},t)$ in Eq. (4) as the Born or single-scattering amplitude, we 
obtain by inverting it

$\hat{\chi}(\hat{s},b) = 2\;i\;\gamma_{\rm{qq}}\;\left (\hat{s}\;e^{-i\frac{\pi}{2}}\right )^{\omega}
      \;K_{0}(\frac{b}{r_{0}})$. \hspace{7.25cm}                                                            (6)\\
Expanding the exponential in (5), we get

$\hat{T}(\hat{s},t) = -i\;\hat{p}\;\hat{W}\;\int_{0}^{\infty}b\;\rm{d}\it{b}\;J_{0}(b q)
     \left [i\;\hat{\chi}-\frac{\hat{\chi}^{2}}{2!}-i\;\frac{\hat{\chi}^{3}}{3!}
     +\;\frac{\hat{\chi}^{4}}{4!}+...\right ]$.  \hspace{2.79cm}                                            (7)\\
The $n^{th}$ term in the series is       

$\hat{T}_{n}(\hat{s},t) = -i\;\frac{(-1)^{n}\;2^{n-1}}{n!}\;
     \gamma^{n}_{\rm{qq}}\;\hat{s}\left (\hat{s}\;e^{-i\frac{\pi}{2}}\right )^{n\omega}
     \int_{0}^{\infty}b\;\rm{d}\it{b}\;J_{0}(b q)\;K_{0}^{n}\left (\frac{b}{r_{0}}\right )$.  \hspace{2.26cm}        (8)\\
Now, $\hat{s} \simeq x x^{\prime} s$, where $x$ and $x^{\prime}$ are the longitudinal momentum 
fractions of the protons carried by the valence quarks (Fig.\ 3).  This leads to the following pp elastic 
amplitude due to qq scattering: 

$T_{\rm{qq}}(s,t) = \hat{T}_{1}(s,t) \; \mathcal{F}^2_1(q_{\perp})\;+\; \hat{T}_{2}(s,t) \;
           \mathcal{F}^2_2(q_{\perp})\;+\;...\;+\;\hat{T}_{n}(s,t) \;\mathcal{F}^2_n(q_{\perp})$; \hspace{1.92cm} (9)\\
$\mathcal{F}_{1},\; \mathcal{F}_{2}\;,...\;\mathcal{F}_{n}\;$ are the structure factors that take into account momentum 
distributions of the valence quarks inside the proton.  Our earlier calculation kept 
only the first term in Eq. (9).  Fig.\ 2 shows that the effect of multiple hard-pomeron exchange 
in pp scattering is to decrease $\rm{d}\sigma/d\it{t}$ at large $\vert t\vert$ compared to 
$\rm{d}\sigma/d\it{t}$ due to single hard-pomeron exchange (dashed line in Fig.\ 2).  For 
$\vert t\vert <$ 3.5 GeV$^2$, there is little effect due to multiple scattering, i.e.\ unitarization.

Results of some of our quantitative calculations are shown in Figs. 4--7.  The solid curve in Fig.\ 4 
represents our calculated total cross section as a function of $\sqrt{s}$.  Dotted curves represent 
the error band given by Cudell et al.\ \cite{Cudell13}.  In Fig.\ 5, solid and dashed curves represent 
our calculated $\rho_{\mathrm{\bar{p}p}}$ and $\rho_{\rm{pp}}$ respectively $(\rho = \mathop{\rm{Re}}\;
T(s,0)/ \mathop{\rm{Im}}\; T(s,0))$.
Dotted curves, as before, represent the error band given by Cudell et al.\  At $\sqrt{s}$=14 TeV, our values of 
$\sigma_{\rm{tot}}$ and $\rho_{\rm{pp}}$ are 109.4 mb and 0.12 respectively.  Fig.\ 6 shows our calculated 
$\rm{d}\sigma/d\it{t}$ for $\mathrm{\bar{p}p}$ elastic scattering at $\sqrt{s}$ = 541 GeV in the Coulomb--hadronic 
interference region using the Kundr\'at--Lokaj\'i$\check{c}$ek formulation (upper curve) and West--Yennie 
formulation (lower curve). Experimental data are from Augier et al.\ \cite{Augier14}.  Fig.\ 7 shows our 
predicted $\rm{d}\sigma/d\it{t}$ for pp elastic scattering at $\sqrt{s}$ = 14 TeV in the Coulomb--hadronic 
interference region.  Finally, in Fig.\ 8, we compare our predicted pp elastic $\rm{d}\sigma/d\it{t}$ at LHC 
with the predictions of other models proposed by various groups: Avila et al.\ \cite{Avila15}, 
Block et al.\ \cite{Block16}, Bourrely et al.\ \cite{Bourrely17}, Desgrolard et al.\ \cite{Desgrolard18}, and 
Petrov et al.\ (three pomeron)\ \cite{Petrov19}.\vspace{-0.2cm}

\flushleft{\bf{Conclusions}}\\
1. Precision measurement of pp elastic $\rm{d}\sigma/d\it{t}$ at LHC by the TOTEM group in the region 
   $\vert t\vert$ = 0 -- 5 GeV$^2$ will be able to distinguish between various proposed models (see Fig. 8).\\
2. In our nucleon-structure model, the qualitative saturation of the Froissart-Martin bound is due 
   to soft hadronic interactions.\\
3. Large $\vert t\vert$ elastic scattering in our model is due to valence quark-quark scattering.  
   This has been described by us as due to the exchange of a hard-pomeron (BFKL pomeron plus 
   next to leading order corrections).\\
4. Unitarization of the hard-pomeron exchange leads to a decrease of $\rm{d}\sigma/d\it{t}$ at 
   large $\vert t\vert$, but has little effect on forward $\rm{d}\sigma/d\it{t}$.\\
5. The nucleon structure that we find embodies salient features of many leading models-- 
   such as Nambu-Jona-Lasinio model, Skyrmion model, nonlinear $\sigma$-model, chiral-bag model-- but,
   at the end, it presents a unique description of the nucleon.\\\vspace{-0.2cm}
\begin{footnotesize}
\bibliographystyle{blois07} 
{\raggedright
\bibliography{blois07}
}
\end{footnotesize}
\begin{figure}[htp]
  \center{
  \includegraphics[height=2.6in]{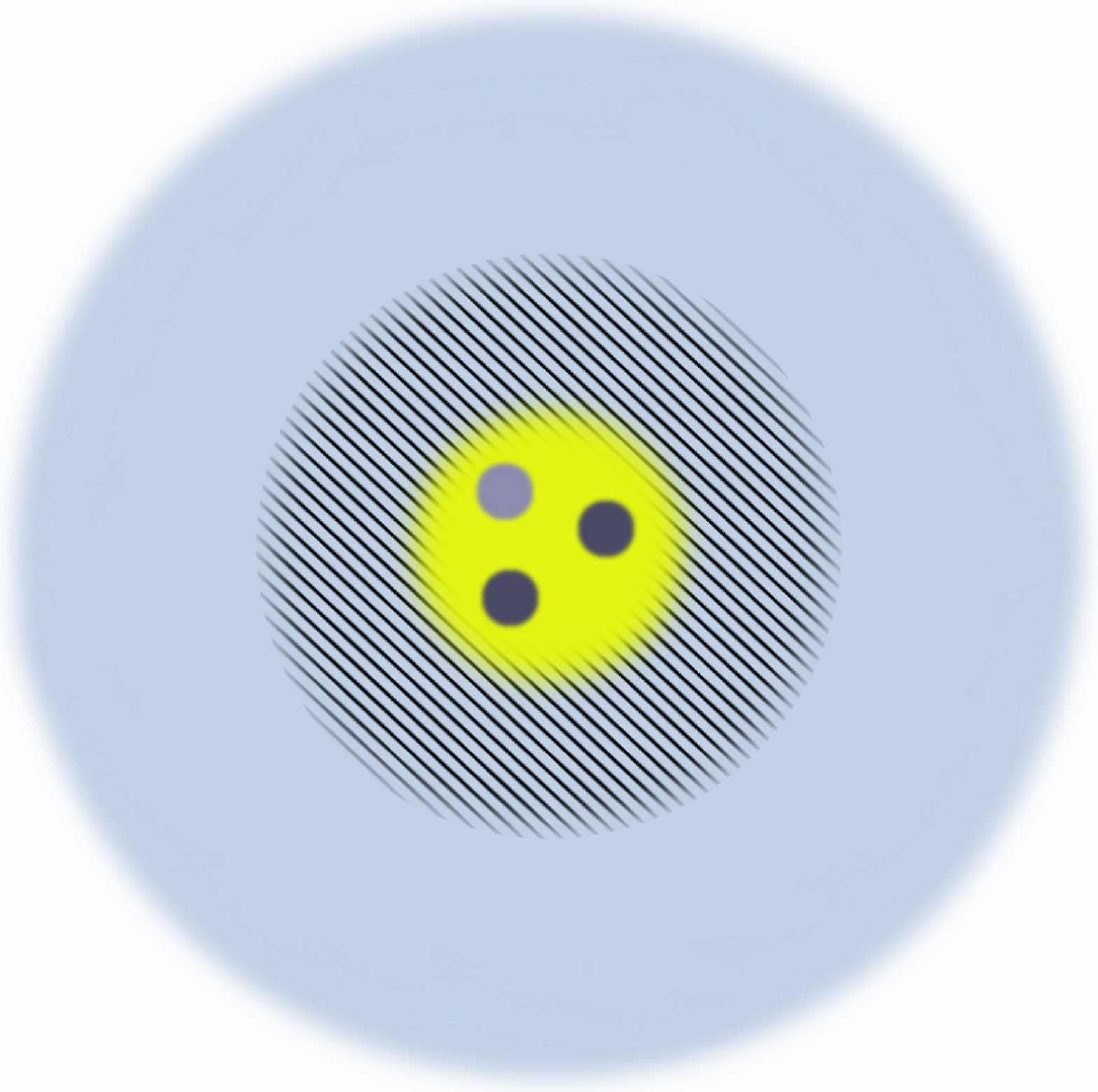}
  \includegraphics[height=2.9in]{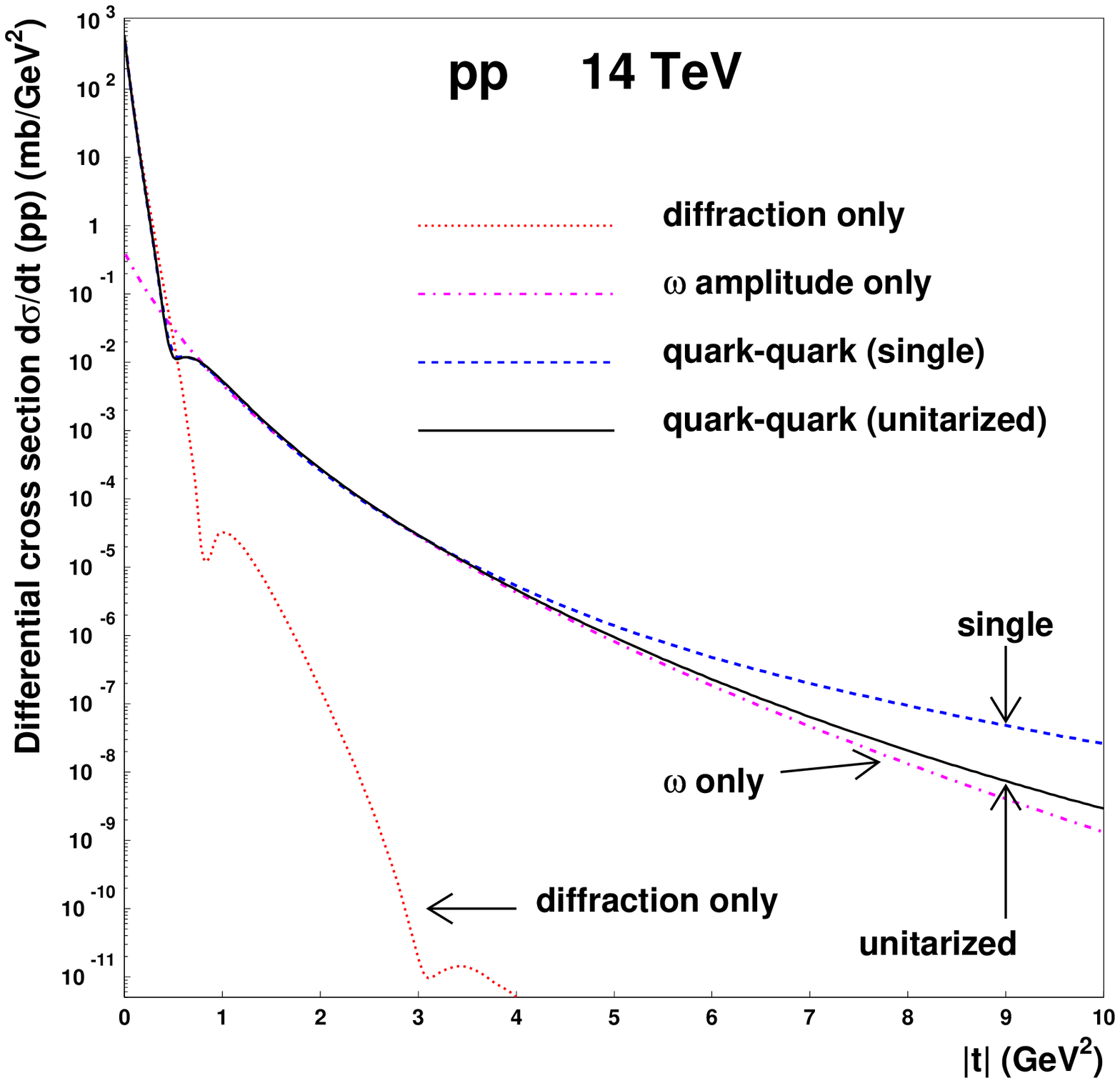}
  \vspace{-0.4cm}
  \caption{\hspace{2.5in}Fig.\ 2}
  }
\end{figure}
\setcounter{figure}{2}
\begin{figure}[htp]
  \vspace{-0.4cm}
  \center{
  \includegraphics[height=1.7in]{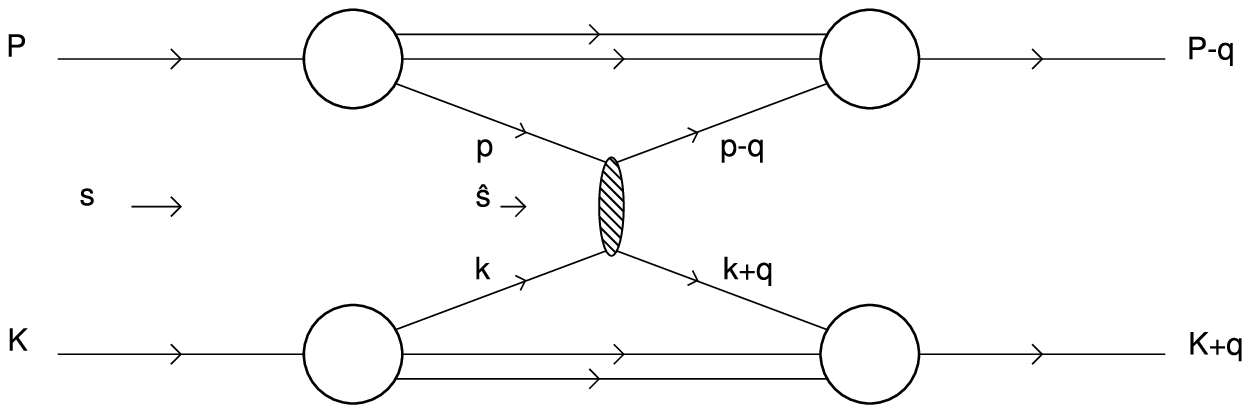}
  \vspace{-0.4cm}
  \caption{}
  }
\end{figure}
\begin{figure}[htp]
\centering
\vspace{-0.8cm}
\includegraphics[height=2.4in]{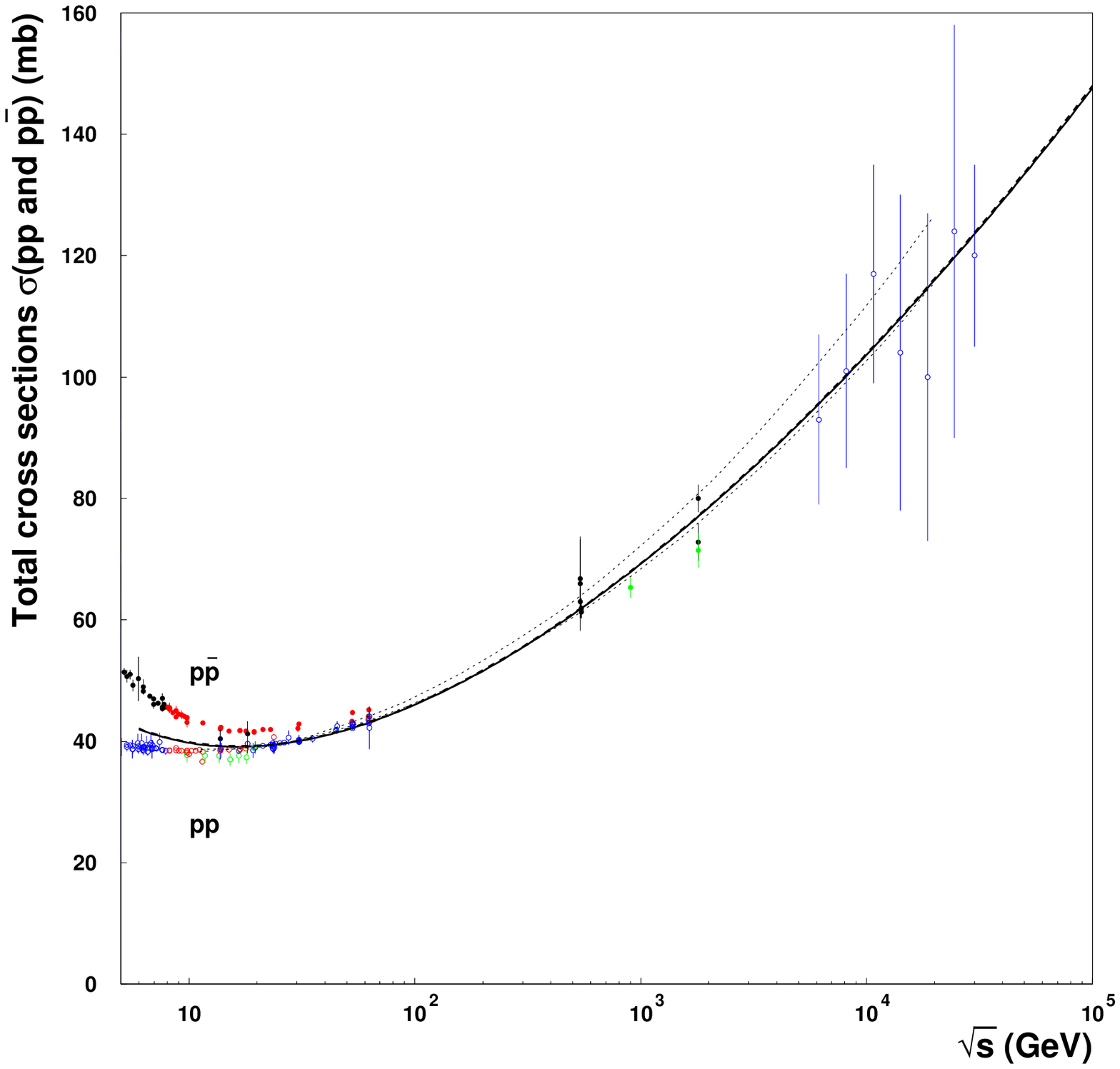}
\hspace{0.1cm}
\includegraphics[height=2.4in]{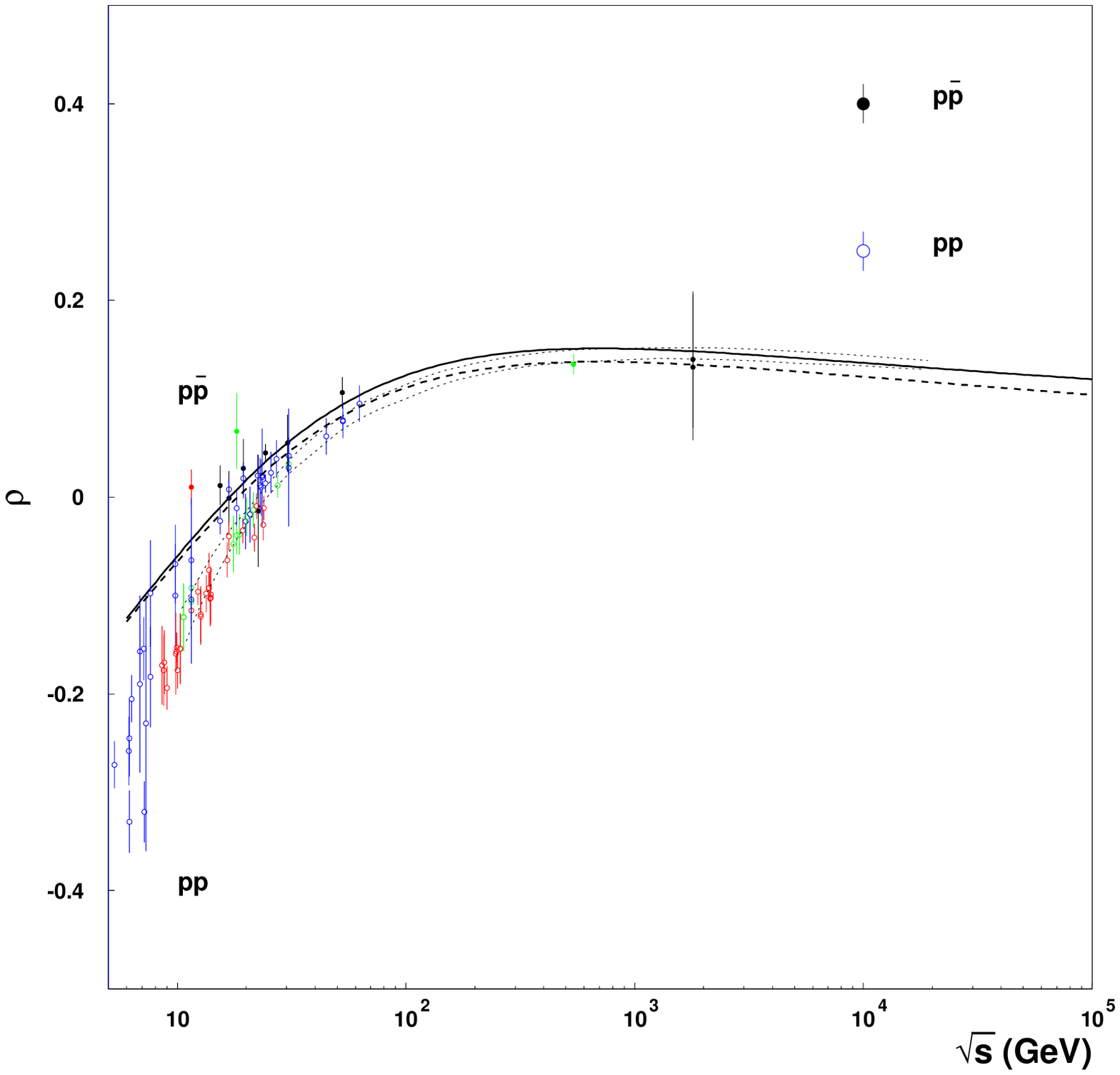}
\vspace{-0.4cm}
\caption{\hspace{6.0cm}Fig.\ 5}
\end{figure}

\setcounter{figure}{5}
\begin{figure}[htp]
\centering
\vspace{-1.0cm}
\includegraphics[height=2.3in]{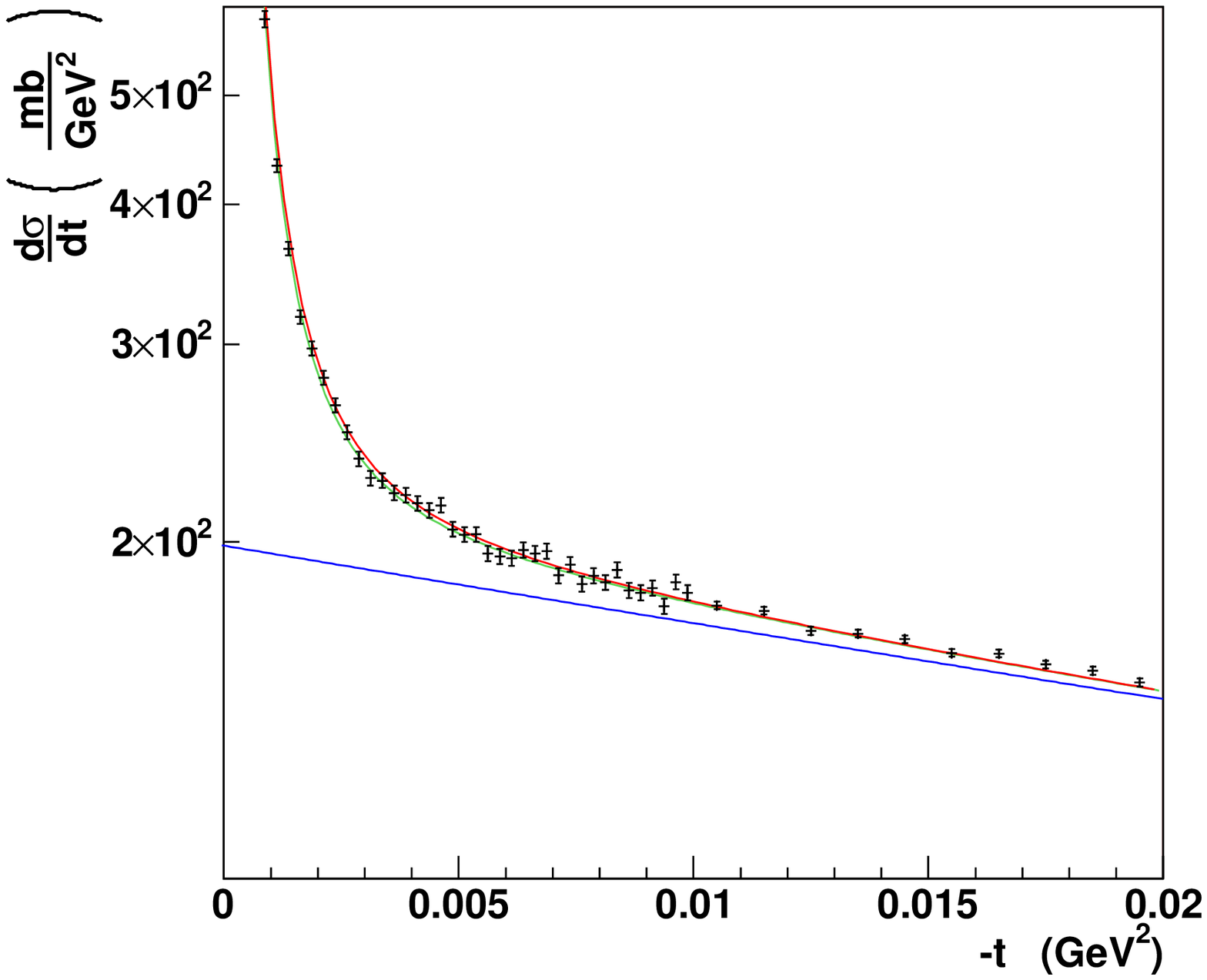}
\hspace{0.0cm}
\includegraphics[height=2.3in]{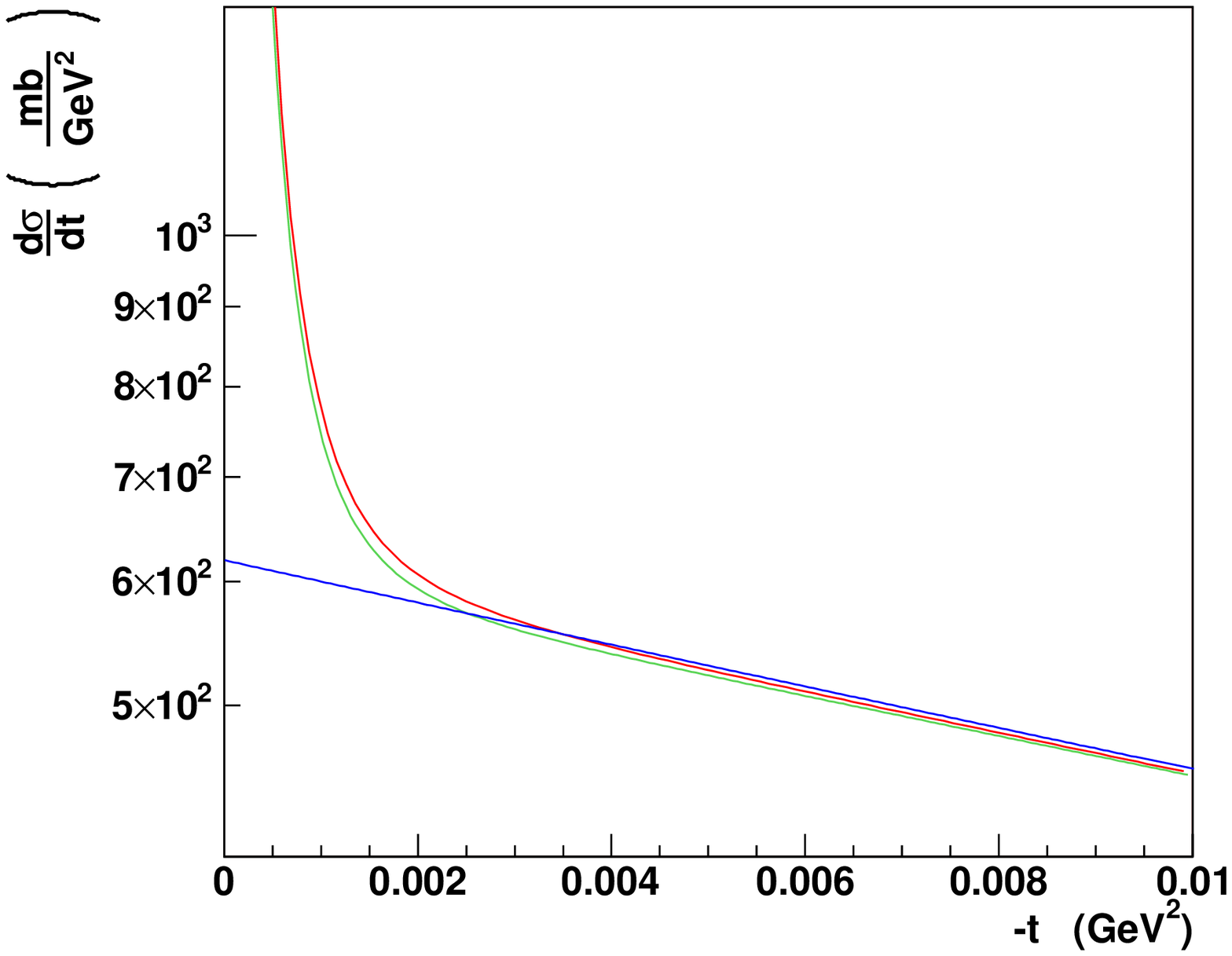} 
\vspace{-0.8cm}
\caption{\hspace{6.0cm}Fig.\ 7}
\end{figure}

\setcounter{figure}{7}
\begin{figure}[htp]
\centering
\vspace{-0.0cm}
\includegraphics[height=5.50in]{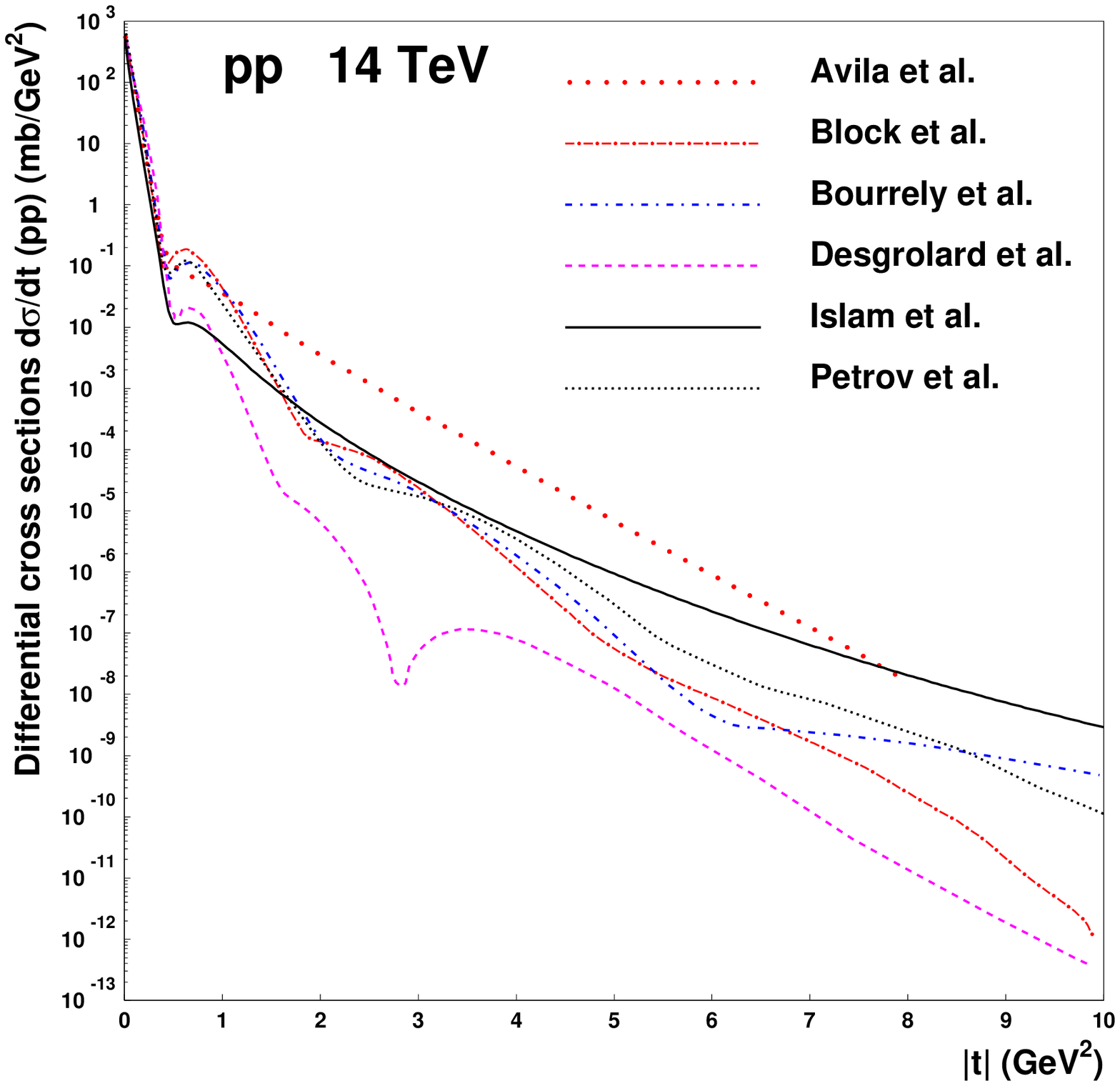} 
\vspace{-1.0cm}
\caption{\hspace{-1cm}   }
\end{figure}

\end{document}